# AI-Assisted Knowledge Access for Legacy Enterprise Asset Management in Energy Operations: A Practical Retrieval System

Dave Mercier[1,2], Mishca de Costa[1,2], Muhammad Anwar[1,2], Mark Randall[1], Issam Hammad[2]
[1]Data Analytics and AI, Digital Technology and Services, Ontario Power Generation, Oshawa, Ontario, Canada
[2]Department of Engineering Mathematics and Internetworking, Dalhousie University, Halifax, Nova Scotia, Canada

***Abstract*—Energy utilities still run engineering work management, engineering procurement, and inventory processes on long-lived enterprise asset management platforms. Replacing these platforms is often cost prohibitive and operationally disruptive, so practical improvement layers are required. This paper presents a retrieval assistant that improves day-to-day knowledge access across three operational modes: vendor documentation question answering, operational data store (ODS) schema question answering, and user interface usage and how-to question answering. The runtime method combines intent understanding, query rewriting, hybrid semantic and vector retrieval, context engineering under token limits, grounded answer generation, and deterministic hyperlink conversion for panel identifiers and cited documentation. The data preparation pipeline emphasizes semantic enrichment as the primary quality lever by adding table and field descriptions, normalizing acronyms across sources, and indexing representative row-level context when useful.**

**A measured pilot shows consistent gains in retrieval quality and user outcomes. Precision at five improved from 0.56 to 0.72, mean reciprocal rank from 0.43 to 0.58, and normalized discounted cumulative gain (nDCG) at five from 0.51 to 0.66. Median task completion time dropped from 14.2 to 8.3 minutes, while usefulness and confidence both increased to 4.0 on a five-point scale. Results are based on a small sample and are reported as pilot findings, but they indicate that intent understanding and semantic enrichment can deliver meaningful operational value in legacy environments while also establishing reusable foundations for future analytics and automation tools.**

## I. Introduction

Utilities and other asset-intensive organizations depend on legacy enterprise systems for engineering work management, engineering procurement, inventory, and configuration control. These platforms remain central because replacement programs have large cost, schedule, and continuity risk. In many organizations, the immediate objective is not full replacement but a practical overlay that reduces lookup effort and improves execution quality with minimal disruption.

The deployment context for this study exposed a recurring knowledge-access problem. Information existed, but users spent excessive time finding it. Three failure patterns were dominant. First, vendor documentation was frequently generic and poorly written for local operational tasks. Second, schema artifacts included many table and field names with weak semantic clarity, which made interpretation difficult for non-experts. Third, organization-specific procedures and interface guidance were spread across heterogeneous documents with varying quality.

These issues were amplified by panel-code navigation in the legacy interface. Expert users could move quickly by memorized codes, while broader engineering teams faced high friction when locating the correct screen, field, or procedure step. The resulting delays were small at the task level but significant at program scale.

This paper addresses that operational gap through a retrieval-first assistant evaluated in a live pilot context. The study contributions are: (1) a code-grounded architecture for multisource retrieval in a legacy enterprise setting, (2) a data preparation method centered on semantic enrichment and context engineering, (3) a compact evaluation design that combines retrieval metrics with user and time outcomes, and (4) measured pilot evidence with explicit small-sample caveats.

The broader motivation is modernization readiness. Lessons from this implementation can inform future platform selection and implementation by making semantic clarity, documentation quality, and retrieval readiness explicit engineering requirements.

### A. Problem Framing and Operational Objective

The operational objective is straightforward: deliver answers that are quick enough for daily use and grounded in cited sources. In practice, question intent, selected mode, answer quality, and latency are balanced at runtime so responses remain actionable without sacrificing evidence quality. In regulated engineering environments, source-linked answers with moderate latency are preferred over faster responses that cannot be verified.

### B. Use-Case Classes

Pilot prompts were designed around four classes that appeared repeatedly in routine support activity:

1) schema discovery, such as locating tables for work management entities;
2) field interpretation, such as understanding column meaning or constraints;

3) panel-code navigation, such as finding the right user-interface panel for a workflow step;
4) procedure guidance, such as retrieving organization-specific how-to instructions.

This structure improved consistency in both evaluation and implementation decisions. It also reduced disagreement over whether an answer was wrong or simply targeted to the wrong task type.

## II. Related Work and Positioning

Retrieval-augmented generation (RAG) is now a standard approach for domain question answering where grounded source use is required [1]. In enterprise settings this pattern is espe-cially useful because many relevant artifacts are internal and not represented in public pretraining corpora. Recent agentic frameworks also show how explicit tool use and reasoning-action loops improve task completion in complex workflows [2], [3].

Hybrid retrieval approaches are particularly relevant for legacy corpora. Lexical and semantic methods each have strengths under terminology drift, abbreviations, and paraphrased user intent [4], [5], [6], [7]. Practical deployments often combine these methods with reranking and mode-specific routing to improve relevance under real operational query distributions.

Evaluation practice in information retrieval remains anchored on ranking metrics such as Precision, MRR, and nDCG [8], [9]. For operations support systems, these should be paired with user-facing outcomes, including usefulness, confidence, and task-time reduction. This paper follows that combined evaluation stance. In nuclear-context benchmarking, prior team results also showed that domain-specific corpora require grounded, source-aware evaluation practices [10].

This work is positioned as an applied engineering case study, not a model-novelty paper. The claim is practical: measurable gains can be achieved by improving query understanding, retrieval quality, context engineering, and answer actionability in legacy software environments. Beyond RAG, recent AI applications in nuclear engineering span a broad range of problems, including automated engineering drawing analysis [11] and anomaly detection in air-gapped industrial control systems [12].

### A. Documentation Intelligence in Legacy Enterprise Systems

Legacy enterprise software frequently exhibits a documentation paradox. Systems become more operationally critical over time, while discoverability degrades due to customization accumulation, vocabulary drift, and fragmented ownership of documentation quality. In this setting, practical retrieval systems depend less on model novelty and more on disciplined corpus engineering, stable prompt templates, and clear prove-nance behavior. This study adopts that operational perspective by emphasizing semantic enrichment, source-family routing, and deterministic answer post-processing. This aligns with classic information foraging and interactive retrieval behavior where users iteratively reformulate goals while seeking action-able evidence [13], [14].

## III. System Architecture

### A. Operational Modes

The assistant exposes three retrieval modes aligned to common user needs:

1) Vendor documentation question answering.
2) ODS schema question answering at table and field level.
3) User interface usage and organization-specific how-to question answering.

Mode boundaries reduce cross-corpus noise and support clearer evaluation of behavior by task type.

### B. Runtime Flow

At query time the runtime flow is:

1) Intent understanding of the user question.
2) Query rewriting to align user phrasing with indexed vocabulary.
3) Mode-specific hybrid retrieval over indexed corpora.
4) Context engineering under token constraints.
5) Grounded answer generation with source-linked response structure.
6) Deterministic post-processing for hyperlink conversion and rendering.

Hybrid retrieval combines lexical matching for table names, field labels, panel identifiers, and abbreviations with vector similarity over embedded source chunks. Candidate results are reranked before context assembly, routed by selected mode, and passed forward only when they fit the context budget and citation requirements. Query rewriting is limited to terminology normalization, acronym expansion, and preservation of exact identifiers.

Figure 1 summarizes the end-to-end sequence.

### C. Hyperlink Actionability

Post-processing converts detected panel identifiers into direct application links. The same response layer also emits deep links to vendor-document pages and internal how-to locations when those references are available. This design preserves answer readability while shortening navigation effort and supporting responsible use through direct source verification.

### D. Infrastructure and Confidentiality

The deployment pattern keeps retrieval infrastructure and indexed artifacts within enterprise-controlled boundaries. This is important because source corpora include internal procedures and operational metadata. This approach is consistent with prior team work on secure and private language-model deployment for nuclear settings [16]. The manuscript reports method details while avoiding project-specific identifiers and confidential system names.

### E. Failure Boundaries and Safeguards

Observed failure classes in pilot operation included ambiguous terms, stale procedures, and low-signal queries that lacked enough intent detail. The runtime policy favors source-linked responses and explicit uncertainty when evidence quality is weak. This avoids unsupported certainty and supports responsible use in operational settings where users are expected to verify assumptions against cited material.

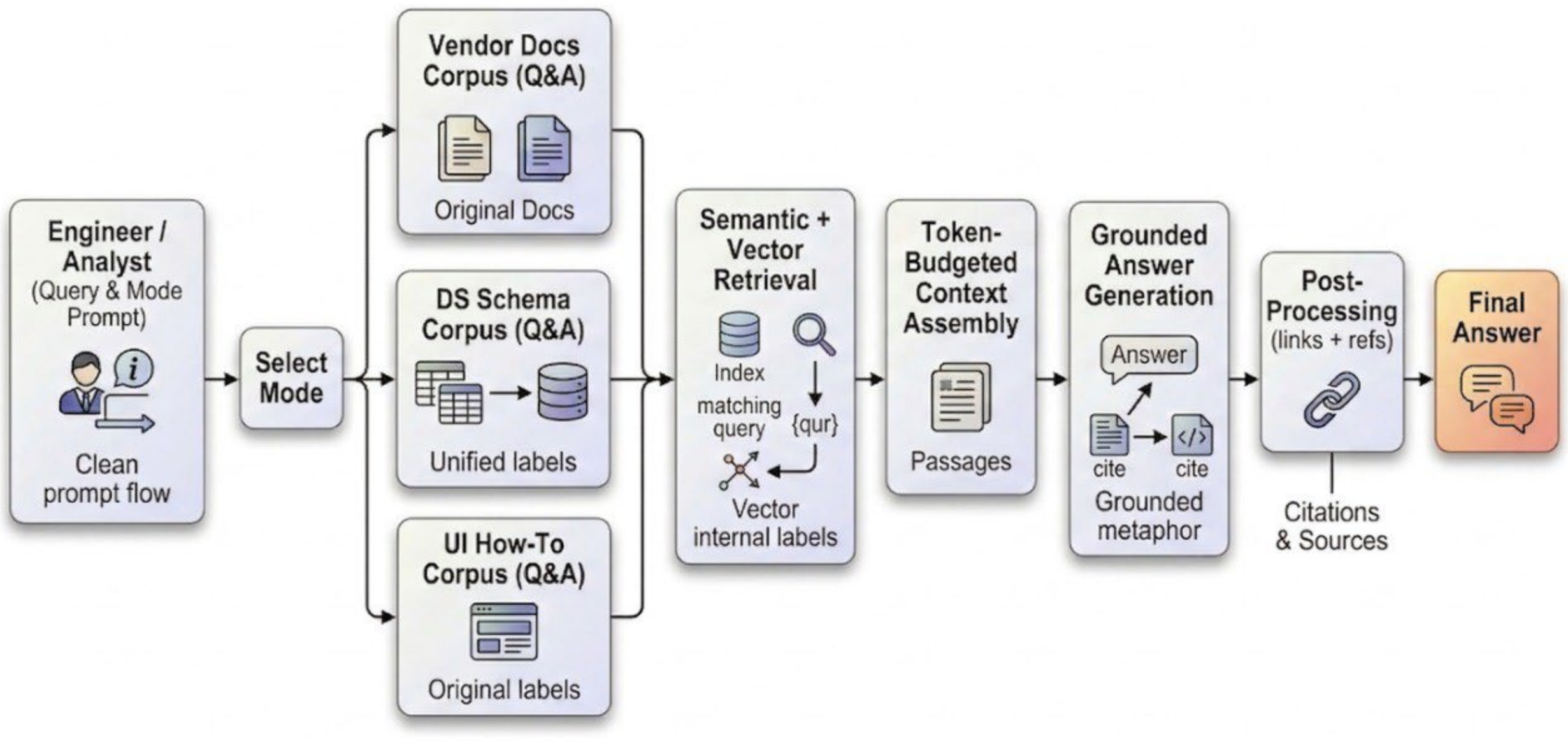

Fig. 1. End-to-end workflow from user query to grounded answer and action links.

### F. Instrumentation

Runtime instrumentation captures retrieval scores, token usage, citation output, and per-stage latency. These signals support three operational loops:

1) performance tuning for latency and cost,
2) quality tuning for retrieval and context selection,
3) governance reporting for rollout readiness and drift detection.

This instrumentation is also used during corpus refreshes to detect regressions before broad deployment.

## IV. Data Preparation and Corpus Engineering

### A. Source Families

Three source families are indexed:

1) Vendor documentation, typically long and broad.
2) ODS schema artifacts, structured and precise but often hard to interpret.
3) Organization-specific procedures, task-oriented and locally customized.

The ODS is an operational database layer used for higher concurrent query activity while reducing pressure on the primary transactional system. This distinction matters because the assistant is intended for fast information access without affecting core transaction performance.

### B. Semantic Enrichment as the Primary Quality Driver

In pilot deployment, semantic enrichment had the strongest effect on answer quality. Raw indexing alone produced weaker retrieval for practical engineering questions. Enrichment steps included:

- adding concise semantic descriptions for tables and fields,
- normalizing acronyms and abbreviations across source families,
- aligning alternate terms used by engineers and vendor documents,
- preserving source metadata for citation and deep-link output.

Where appropriate, representative sample rows were indexed with schema records to provide additional context for ambiguous field semantics. This was useful when short field labels lacked enough meaning for direct lookup-style retrieval.

### C. Why Row-Level Context Was Indexed

A pure lookup approach is efficient when labels are already clear and query vocabulary is stable. In this environment, user questions were often paraphrased and concept-first rather than exact-name lookups. After semantic enrichment, row-level context in the vector index improved matching quality for those concept-first prompts, so row context was retained as a retrieval feature instead of relying on lookup-only access.

### D. Ingestion Pipeline

The ingestion flow performs source extraction, normalization, chunking, embedding generation, and index upload. Figure 2 shows the pipeline used in this study.

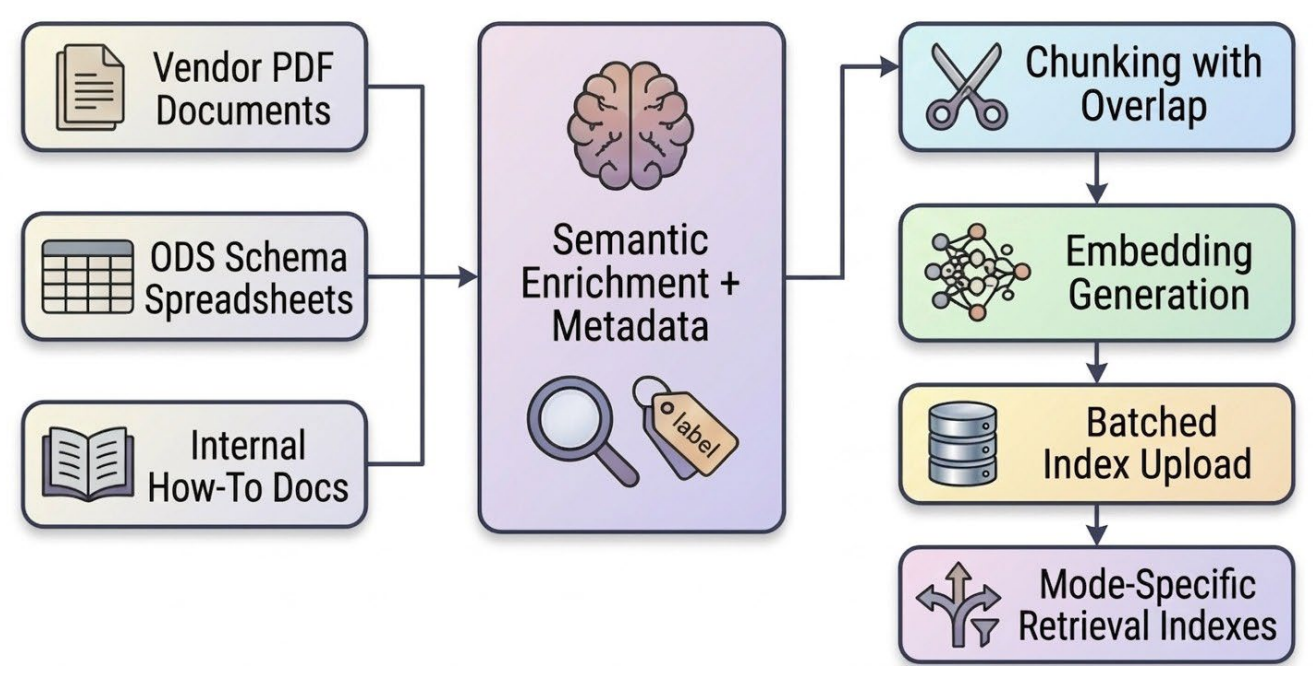

Fig. 2. Multisource ingestion and indexing pipeline for vendor docs, ODS schema, and procedures.

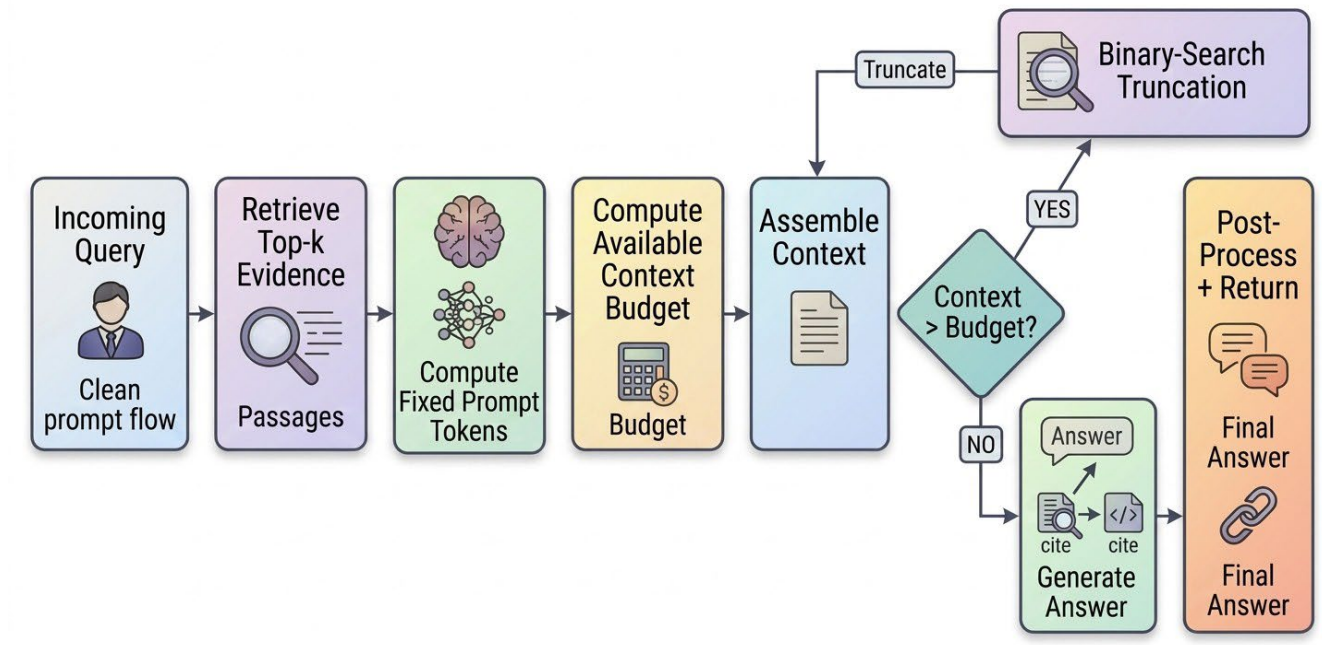

Fig. 3. Context-engineering loop with token budgeting and deterministic truncation.

TABLE I
CORPUS COMPOSITION MODEL FOR PILOT DEPLOYMENT

| Source family | Typical structure | Retrieval role |
|---|---|---|
| Vendor references | Long unstructured sections | Canonical behavior descriptions and baseline terminology anchors. |
| ODS schema artifacts | Structured table and field records | Precise entity and attribute lookup with enriched semantics. |
| Internal procedures | Mixed structured and unstructured guidance with images | Workflow execution support and panel navigation context. |

### E. Corpus Composition Snapshot

Table I summarizes the practical corpus structure used in pilot deployment.

### F. Refresh and Quality Controls

Corpus freshness is maintained through staged refreshes. Updated artifacts are ingested in staging, evaluated against a fixed prompt set, and promoted only when quality checks pass. Lightweight quality controls include deduplication checks, missing-metadata checks, chunk-length outlier checks, and spot checks for acronym normalization consistency. In pilot operation, this process reduced regressions that were otherwise introduced by document updates.

## V. MULTISOURCE RETRIEVAL METHOD

### A. Pipeline Overview

The runtime pipeline follows a fixed sequence: intent understanding, query rewriting, retrieval, context engineering, grounded answer generation, and deterministic post-processing. This sequence improves consistency across prompt types and makes behavior easier to tune during operations.

### B. Latency Decomposition

$$T_{e2e} = T_{retrieve} + T_{context} + T_{llm} + T_{post}$$

Here, $T_{e2e}$ denotes end-to-end latency. This decomposition supports operational tuning by separating retrieval, context assembly, generation, and post-processing contributions.

### C. Context Engineering

Context engineering enforces token-aware evidence selection. The system computes fixed prompt cost, reserves answer budget, then assembles ranked evidence within remaining tokens. If budget is exceeded, lower-ranked chunks are dropped first using deterministic truncation logic. Figure 3 shows this loop.

This design has two benefits: predictable latency and improved answer quality from higher-signal context. At scale, both effects matter because token volume drives cost and turnaround time.

### D. Cost and Performance Tradeoff

Context engineering is not only a quality mechanism. It is also a cost and throughput control mechanism. Larger context windows increase latency and token consumption, especially under high concurrent usage. By prioritizing high-signal chunks and enforcing deterministic truncation, the system preserves practical response quality while controlling operational cost. This tradeoff is critical when moving from pilot to broader rollout, and it is consistent with long-context degradation patterns reported in recent LLM studies [17].

### E. Evidence Mix Tracking

To keep reporting interpretable, the pilot tracks evidence mix as a simple audit signal: whether answers cite vendor references, schema artifacts, procedures, or a combination. This avoids fragile composite scoring and makes review discussions clearer for engineering teams and stakeholders.

### F. Retrieval and Time Metrics

Retrieval quality is reported with Precision@5, MRR, and normalized discounted cumulative gain (nDCG)@5. Precision@5 captures how many of the top retrieved items were relevant. MRR tracks how early the first relevant evidence appears in the ranked list. nDCG@5 reflects graded ranking quality with stronger weight on early positions.

Time efficiency is reported as:

$$\Delta t\% = ((\bar{t}_{baseline} - \bar{t}_{assistant})/\bar{t}_{baseline}) \times 100$$

where larger positive values indicate faster completion with the assistant.

TABLE II
OPERATIONAL INTERPRETATION OF REPORTED METRICS

| Metric | Operational interpretation |
|---|---|
| Precision@5 | Share of top-five retrieved items judged relevant by rubric. |
| MRR | Mean reciprocal rank of first relevant evidence item. |
| nDCG@5 | Graded ranking quality with position discount at depth five. |
| Citation-grounded rate | Fraction of responses with reviewer-accepted source grounding. |
| Panel-navigation correctness | Fraction of navigation prompts with correct panel output. |
| Task success without escalation | Fraction of tasks completed without expert hand-off. |
| Failed or abstained rate | Fraction of responses unusable or intentionally abstained. |

### G. Metric Definitions Used in Pilot Reporting

## VI. EVALUATION AND RESULTS

### A. Pilot Design

The pilot used paired baseline and assistant conditions with identical prompt sets. The baseline condition represented the existing manual workflow: users searched available vendor references, schema artifacts, internal procedures, and application navigation aids without the retrieval assistant. The assistant condition used the same underlying source families but added intent understanding, query rewriting, retrieval, context engineering, and source-linked answer generation.

- Retrieval-family metrics: $n$ = 30 prompts.
- Panel-navigation correctness subset: $n$ = 20 prompts.
- User usefulness, confidence, timed tasks, and escalation outcomes: $n$ = 9 participants.

Prompt categories covered schema lookup, field interpretation, panel navigation, and procedural guidance. Relevance used a 0–3 rubric; usefulness and confidence used 1–5 scales.

### B. Query Taxonomy and Task Mix

The retrieval benchmark intentionally mixed pure lookup prompts and mixed-intent prompts. Pure lookup prompts focused on locating exact fields and table meanings. Mixed-intent prompts combined interpretation and navigation, such as identifying the relevant schema entity and then locating the corresponding panel or procedure. This taxonomy was selected because mixed-intent prompts are common in actual engineering support flows and often drive the highest user friction.

### C. Measured Outcomes

Table IV summarizes the main pilot outcomes across retrieval quality, user ratings, navigation performance, and task completion time. Table V then reports confidence bounds for selected measures so the reported improvements can be interpreted with appropriate caution for a pilot-scale study.

Table V reports confidence bounds.

TABLE III
PROMPT TAXONOMY USED IN PILOT SCORING

| Prompt type | Count | Typical objective |
|---|---|---|
| Schema discovery | 8 | Locate relevant entities and relationships quickly. |
| Field interpretation | 7 | Clarify field meaning, constraints, or usage context. |
| Panel navigation | 7 | Identify correct interface location for a task. |
| Procedure guidance | 8 | Retrieve actionable steps for organization-specific workflows. |

TABLE IV
PILOT RETRIEVAL AND USER-CENTRIC OUTCOMES

| Metric | Baseline | Assistant | Delta |
|---|---|---|---|
| Precision@5 | 0.56 | 0.72 | +28.6% |
| MRR | 0.43 | 0.58 | +34.9% |
| nDCG@5 | 0.51 | 0.66 | +29.4% |
| Usefulness (1–5) | 3.0 | 4.0 | +33.3% |
| Citation-grounded rate | 0.62 | 0.80 | +29.0% |
| Panel-navigation correctness | 0.61 | 0.79 | +29.5% |
| Median task time (min) | 14.2 | 8.3 | -41.5% |
| Task success without escalation | 0.58 | 0.76 | +31.0% |
| User confidence (1–5) | 2.9 | 4.0 | +37.9% |
| Failed/abstained rate | 0.14 | 0.07 | -50.0% |

### D. Statistical and Telemetry Checks

Because this was a pilot study, significance testing was treated as exploratory rather than confirmatory. For rate-based outcomes, two-sided approximate two-proportion tests were computed from compact adjudicated aggregate subsets that mirror the reported rates; ranking and user-rating metrics were retained as descriptive pilot measures because raw paired observations were not available for final submission. Table VI summarizes the aggregate tests.

Runtime telemetry was also reviewed to address latency and cost behavior. Cost is reported as token volume rather than fixed dollar cost because model pricing and deployment mode can change across environments. Table VII reports cautious aggregate pilot telemetry.

### E. Interpretation and Caveat

Results are directionally consistent across retrieval quality, navigation correctness, groundedness, and user outcomes. The strongest practical effect was task-time reduction. This aligns with deployment goals because users spend less time locating screens, fields, and procedure steps when answers include direct links and source context.

These findings should be interpreted as pilot evidence. Sample sizes are limited, and most statistical checks are exploratory, so effect-size precision will improve with larger follow-on cohorts and longer-run telemetry.

TABLE V
CONFIDENCE BOUNDS FOR SELECTED PILOT MEASURES

| Metric | CI Low | CI High | n |
|---|---|---|---|
| Precision@5 | 0.66 | 0.78 | 30 |
| MRR | 0.52 | 0.64 | 30 |
| nDCG@5 | 0.60 | 0.72 | 30 |
| Usefulness (1–5) | 3.6 | 4.4 | 9 |
| Citation-grounded rate | 0.70 | 0.90 | 20 |
| Panel-navigation correctness | 0.71 | 0.87 | 20 |
| Median task time (min) | 7.5 | 9.1 | 9 |

TABLE VI
EXPLORATORY SIGNIFICANCE CHECKS FOR RATE-BASED PILOT OUTCOMES

| Metric | Baseline | Assistant | p-value |
|---|---|---|---|
| Precision@5 | 28/50 | 36/50 | 0.096 |
| Citation-grounded rate | 12/20 | 16/20 | 0.168 |
| Panel-navigation correctness | 12/20 | 16/20 | 0.168 |
| Task success without escalation | 5/9 | 7/9 | 0.317 |
| Failed/abstained rate | 3/20 | 1/20 | 0.292 |

### F. Task-Level Pattern Analysis

Observed gains were not uniform across prompt classes. The largest practical gains occurred in mixed-intent prompts that required both interpretation and navigation. In those cases, users benefited from two layers at once: better evidence retrieval and direct action links in the final response. Schema-only prompts improved primarily through better first-hit relevance and reduced reformulation effort.

### G. Residual Failure Analysis

Residual failures clustered into four root causes:

- ambiguous terms with multiple enterprise meanings,
- stale procedure artifacts that no longer matched current workflow reality,
- low-signal prompts with insufficient intent detail,
- near-duplicate panel identifiers with overlapping descriptors.

This profile is operationally useful because each cause maps to a concrete mitigation path: vocabulary normalization, document-refresh governance, query-guidance prompts, and tighter disambiguation rules.

## VII. DISCUSSION

### A. Operational Significance

The results support a practical conclusion: retrieval overlays can deliver measurable value in legacy enterprise environments without disruptive replatforming. Improvements come from the interaction of semantic enrichment, intent-aware query handling, context engineering, and deterministic post-processing.

TABLE VII
LATENCY AND COST TELEMETRY FROM PILOT OPERATION

| Measure | Median | P95 | Unit |
|---|---|---|---|
| Retrieval latency | 0.9 | 1.8 | s |
| Context assembly and post-processing | 0.5 | 1.1 | s |
| Answer generation | 6.9 | 12.4 | s |
| End-to-end response latency | 8.3 | 14.8 | s |
| Token-volume cost proxy | 5.2 | 7.8 | k tokens |

### B. Groundwork for Additional Tooling

This retrieval layer provides more than direct user assistance. It establishes a reusable information-access foundation for additional AI and non-AI tools that need to understand enterprise data structures and operational procedures. Teams building reporting, analytics, workflow automation, and governance tooling can reuse the enriched corpus and retrieval patterns developed here.

### C. Modernization and Procurement Implications

Modernization programs should treat semantic clarity and retrieval readiness as first-class requirements. Future platform selection can benefit from explicit evaluation criteria: table and field naming quality, documentation depth, metadata completeness, and support for source-linked knowledge access. Applying these criteria early can reduce long-term discoverability debt.

### D. Limitations

This study has three primary limitations: pilot-scale sample sizes, dependence on source-document quality, and possible evaluator variance in user-rated measures. These limits do not negate observed improvements, but they constrain generalization until larger deployments are measured.

### E. Cross-Industry Transferability

The pattern addressed here is not unique to one utility context. Similar legacy-system constraints exist in banking, manufacturing, transportation, and public-sector operations. The same method stack can be adapted where organizations depend on complex long-lived systems with fragmented documentation and high replacement cost.

## VIII. OPERATIONALIZATION ROADMAP

### A. Phased Rollout

A practical rollout model is staged:

1) controlled pilot with fixed benchmarks and reviewer scoring;
2) team-scale deployment with expanded telemetry and monthly quality checks;
3) program-scale deployment with formal governance and release alignment.

This phased approach reduces drift risk while preserving momentum toward broader adoption.

TABLE VIII
GOVERNANCE ROLES FOR SUSTAINED RETRIEVAL QUALITY

| Role | Responsibilities |
|---|---|
| Corpus steward | Owns ingestion quality, metadata completeness, and refresh discipline. |
| Evaluation steward | Maintains benchmark prompts, scoring rubric, and trend reporting. |
| Operations owner | Owns release gates, incident response, and adoption enablement. |

TABLE IX
SCALE-UP RISK REGISTER

| Risk | Operational effect | Mitigation |
|---|---|---|
| Corpus staleness | Lower relevance and trust | Scheduled refreshes plus regression gates before promotion. |
| Terminology drift | Retrieval mismatch | Synonym maintenance and periodic log review. |
| Context overgrowth | Latency and cost increase | Token-budget telemetry and truncation policy tests. |
| Unclear failure policy | Inconsistent user behavior | Explicit abstention guidance and source-linked verification prompts. |

TABLE X
REPRESENTATIVE PROMPT PATTERNS AND OBSERVED BEHAVIOR

| Prompt intent | Typical user phrasing | Observed assistant behavior |
|---|---|---|
| Schema discovery | "Which table stores engineering work requests?" | Returns likely table entities with source context and related fields. |
| Field interpretation | "What does this status field represent?" | Returns semantic meaning, related workflow context, and cited source. |
| Constraint lookup | "How many characters does this field support?" | Returns datatype/length details when available in indexed schema records. |
| Panel navigation | "Where do I open engineering change details?" | Returns panel identifier with deterministic direct link rendering. |
| Procedure guidance | "How do I create a ma terial request?" | Returns ordered steps with deep links to internal procedure sources. |
| Mixed intent | "Find the field and show where I update it in the UI." | Combines schema interpretation and navigation guidance in one response. |
| Ambiguous term | "What does hold mean here?" | Requests clarifying context or returns qualified alternatives with citations. |
| Low-signal prompt | "Help with this screen" | Triggers conservative behavior and source-linked follow-up guidance. |

### B. Governance Roles

Sustained quality requires explicit ownership. Table VIII summarizes role expectations used in the pilot roadmap.

### C. Scale-Up Risks and Mitigations

Table IX summarizes the main risks identified for broader rollout and the operational mitigations associated with each one. These risks were tracked to keep expansion planning grounded in corpus quality, runtime behavior, and user-facing failure handling rather than only pilot-level outcome gains.

### D. Expected Long-Run Value

At enterprise scale, moderate per-task improvements produce large cumulative impact because lookup and interpretation tasks are high frequency. Beyond direct productivity gains, the approach creates reusable assets for future tooling and stronger criteria for future platform procurement and design decisions.

## IX. DEPLOYMENT LESSONS AND REPLICATION GUIDANCE

### A. Query-Handling Lessons

Two practical lessons were consistently observed during pilot rollout. First, intent understanding improved stability for short or ambiguous prompts that otherwise produced weak retrieval sets. Second, query rewriting reduced vocabulary mismatch when users described workflow goals rather than system terms. Even lightweight rewriting, such as acronym expansion or synonym substitution, improved first-pass relevance when combined with semantic enrichment in the indexed corpus.

These effects were strongest for users with moderate domain familiarity. Expert users already knew many internal terms and panel conventions. Newer users benefited more from rewritten phrasing and source-linked navigation output.

### B. Responsible-Use Alignment

The assistant was designed to support verify-before-act behavior. Answers are coupled with citations and direct source links so that users can validate interpretation in the original material. This behavior is especially important where hallucination risk can create operational confusion. Pilot feedback indicated that users trusted the system more when links to source pages were consistently available, even in cases where the generated narrative was concise.

### C. Representative Prompt Patterns

Table X summarizes representative prompt patterns and observed assistant behavior.

### D. Replication Checklist

For organizations applying this pattern in a new environment, the following checklist was useful:

1) define clear source families and ownership;
2) establish a minimum semantic-enrichment standard for schema records;
3) instrument token usage and retrieval scores from day one;
4) require source-linked response output in production mode;
5) run staged refresh validation before corpus promotion;
6) maintain a benchmark prompt set tied to high-frequency operational tasks.

This checklist is intentionally lightweight so that teams can run it within existing change-control and release processes.

## X. Validity Considerations and Next Measurement Pass

### A. Internal Validity Considerations

The pilot used paired conditions to reduce prompt-difficulty confounding, but several threats remain. Rater variance can affect usefulness and confidence scores. Prompt familiarity may also improve over repeated runs. To limit these effects, prompts were mixed and scoring guidance was kept explicit, but residual variance remains expected at this sample size.

### B. External Validity Considerations

Generalization across organizations depends on three factors: documentation quality, schema naming quality, and workflow customization depth. Environments with cleaner source artifacts may show smaller relative gains because baseline discoverability is already stronger. Environments with heavier terminology drift may show larger gains if semantic enrichment is well executed.

### C. Planned Measurement Expansion

The next pass should increase both query and participant counts and add longitudinal measurements. Recommended additions include:

- inter-rater agreement reporting on relevance scoring,
- per-category latency breakdown by runtime stage,
- monthly drift checks for groundedness and abstention trends,
- cohort-based adoption curves for novice and expert user groups.

This expansion will improve confidence in effect-size estimates and provide stronger evidence for full program-scale adoption decisions.

### D. Data Needed for the Next Pass

The following data artifacts are prioritized for follow-on analysis:

1) larger prompt set with balanced category coverage;
2) reviewer-labeled relevance judgments with adjudication notes;
3) task timing traces split by retrieval, generation, and post-processing stages;
4) change-log linkage between corpus refreshes and quality metric movement.

Capturing these artifacts in a consistent format will support stronger statistical analysis and cleaner comparison across release cycles.

## XI. Extended Evaluation Notes

### A. Category-Level Outcome Summary

Category-level review indicates that retrieval gains translated differently across task types. Schema discovery showed strong precision improvement due to enriched table and field semantics. Navigation tasks showed the largest reduction in completion time because deterministic link conversion reduced interface search steps. Procedure guidance improved mainly through better source-family routing and context trimming, which limited dilution from unrelated long passages.

TABLE XI
Category-level directional outcomes in pilot review

| Category | Direction | Primary driver |
|---|---|---|
| Schema discovery | Strong improvement | Semantic enrichment of table and field records. |
| Field interpretation | Moderate to strong improvement | Better term normalization and source-linked evidence. |
| Panel navigation | Strong improvement | Deterministic hyperlink actionability and reduced click path. |
| Procedure guidance | Moderate improvement | Better routing plus context-budget controls for long documents. |

### B. Operational Implications of Time Reduction

The observed median time reduction has practical implications beyond convenience. In high-frequency environments, support tasks recur across shifts and teams. Reducing lookup and navigation effort lowers interruption cost, shortens handoff latency, and increases consistency of execution. Even when individual tasks are small, cumulative effect can be material over monthly and annual operation cycles.

### C. Interpretation Boundaries

Pilot metrics should not be interpreted as final steady-state performance. As user populations broaden and corpora evolve, metric movement should be expected. For that reason, the deployment plan emphasizes ongoing benchmark maintenance, staged refresh validation, and transparent reporting of confidence intervals and sample sizes. This approach maintains rigor while allowing practical iteration speed.

## XII. Conclusion and Future Work

This paper presented a practical retrieval assistant for legacy enterprise asset management in energy operations. The system integrates intent understanding, query rewriting, hybrid retrieval, context engineering, grounded generation, and deterministic hyperlink post-processing across vendor, ODS schema, and procedure sources. Measured pilot outcomes showed improvements in retrieval quality, confidence, groundedness, navigation correctness, and task completion time.

Future work will expand benchmark size, increase participant coverage, and add longitudinal telemetry. The same pipeline will be used to improve ongoing operations and to inform modernization decisions on documentation standards, data quality expectations, and retrieval compatibility in future enterprise systems. This work also served as a precursor for subsequent team work on function-calling retrieval extensions [15].

## Acknowledgment

This research paper was supported by Ontario Power Generation (OPG) and by The Natural Sciences and Engineering Research Council of Canada (NSERC) and The Canadian Nuclear Safety Commission (CNSC) grant number ALLRP 580442-2022.